\documentclass[%
reprint,
showpacs,preprintnumbers,
amsmath,amssymb,
longbibliography
]{revtex4-2}

\usepackage{graphicx}
\usepackage{dcolumn}
\usepackage{bm}
\usepackage{colortbl}
\usepackage[unicode=true,colorlinks=true,citecolor=blue,urlcolor=blue]{hyperref}
\usepackage{braket}

\graphicspath{{img/}}

\newcommand{\e}{\mathrm{e}}

\let\ifr\i
\renewcommand{\i}{{\rm i}}
\renewcommand{\d}{\mathrm d}
\renewcommand{\emph}{\textit}

\newcommand{\addDima}[1]{#1}

\begin{document}

\title{Tunneling spin Nernst effect for a single quantum dot}

\author{V.~N.~Mantsevich}
\affiliation{Chair of Semiconductors and Cryoelectronics, Faculty of Physics, Lomonosov Moscow State University, 119991 Moscow, Russia}
\author{D.~S.~Smirnov}
\email[Electronic address: ]{smirnov@mail.ioffe.ru}
\affiliation{Ioffe Institute, 194021 St. Petersburg, Russia}

\begin{abstract}
  We describe theoretically the spin Nernst effect for electrons tunneling to a quantum dot from a quantum wire with the heat flowing along it. Such a tunneling spin Nernst effect is shown to take place due to the spin-dependent electron tunneling produced by the spin-orbit coupling. The Coulomb interaction of electrons in the quantum dot is taken into account using nonequilibrium Green's functions and is shown to increase significantly the accumulated spin in a single quantum dot. The largest possible degree of spin polarization is discussed.
\end{abstract}

\maketitle{}

\section{Introduction}

The spin Nernst effect is a hybrid of the Nernst and spin Hall effects. The former represents generation of the transverse electric current under the longitudinal flow of heat. The latter is the transverse spin current induced by the longitudinal electric current. Thus the spin Nernst effect describes the appearance of the transverse spin current in response to the longitudinal heat flow~\cite{bose19}.

This effect was first predicted in the external magnetic field~\cite{cheng08}, but later it was realized that the magnetic field is not necessary~\cite{liu10} as for the spin Hall effect~\cite{dyakonov71}. These pioneering works were followed by first principles~\cite{tauber12,wimmer13,guo17} and kinetic~\cite{kovalev16,PhysRevB.106.045203} calculations for various systems. Despite early attempts to observe the spin Nernst effect~\cite{seki10}, the conclusive experiments were performed only recently~\cite{meyer17,sheng2017,kim2017}.

This effect was never observed in semiconductors (to the best of our knowledge) due to the weakness of spin-orbit coupling~\cite{ganichev2012spin}. So it is important to propose a setup that will help to demonstrate appearance of spin polarization without electric current, only with the heat flow. The electron localization in quantum dots, for example, can be exploited to increase the spin relaxation time and the steady state spin polarization~\cite{Hopping_spin}. In addition, the electron spin accumulated in a quantum dot, can be conveniently measured and manipulated by various optical and electrical means~\cite{book_Glazov}.

The spin accumulation in localized states requires tunneling which depends on the electron spin. Thus in this work we will be concerned with the \emph{tunneling spin Nernst effect}. The spin dependent tunneling was shown to take place due to the Dresselhaus and Rashba spin-orbit interactions~\cite{perel:201304,tarasenko:056601,Tarasenko_Tunneling}, so the effect takes place in nonmagnetic structures. In recent works, large current induced spin accumulation was predicted for the hopping conductivity regime~\cite{Hopping_spin,PhysRevB.98.155304} and for the tunneling of holes~\cite{Mantsevich2022} and electrons~\cite{Mantsevich2023} to a single quantum dot. Thus a large spin accumulation due to the tunneling spin Nernst effect may be expected.

In this paper, we describe the tunneling spin Nernst effect for a single quantum dot side coupled to a ballistic quantum wire. The system and the method of nonequilibrium Green's functions are described in Sec.~\ref{sec:model}. The results of the calculation of the spin accumulation induced by the heat current are presented in Sec.~\ref{sec:results}, where we show that the possible degree of spin polarization can be large indeed. Sec.~\ref{sec:conclusion} concludes the paper.

\section{Model}
\label{sec:model}

\begin{figure}
\includegraphics[width=0.75\linewidth]{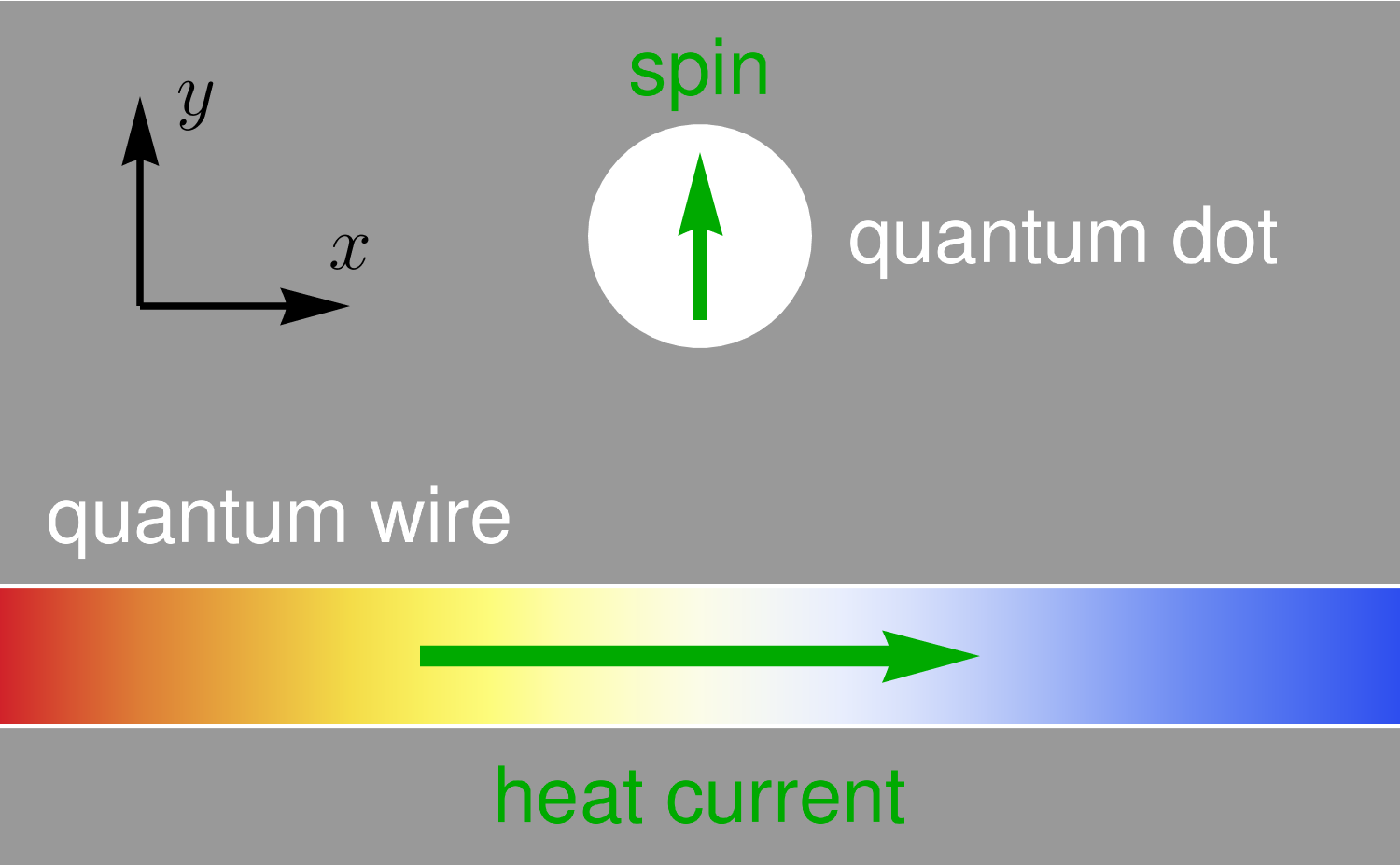}
\caption{Scheme of the system under investigation. A heat current flows along the quantum wire due to the temperature gradient between the ends and leads to the spin accumulation in the quantum dot. }
\label{fig:system}
\end{figure}

We study the spin accumulation in the quantum dot (QD) induced by the heat flow in the quantum wire due to the temperature gradient between its ends, see Fig.~\ref{fig:system}. The system is assumed to be gate defined in a two dimensional electron gas~\cite{exp1,exp2,exp3}, but the other realizations are possible as well~\cite{review1}. The $C_{2v}$ symmetry group of the system allows for the linear coupling between the temperature gradient along the wire and the spin polarization in the QD along the structure growth axis.

To describe this coupling microscopically, we use the following Anderson-like Hamiltonian~\cite{PhysRev.124.41}:
\begin{multline}
  \label{eq:Ham_m}
  \mathcal H=E_0\sum_\sigma n_\sigma+Un_{\sigma}n_{-\sigma}+\sum_{k,\sigma}E_kn_{k,\sigma}\\
  +\sum_{k,\sigma}\left(V_{k,\sigma}d_\sigma^\dag c_{k,\sigma}+{\rm H.c.}\right).
\end{multline}
Here $E_0$ is a single electron energy in the QD, $n_\sigma=d_\sigma^\dag d_\sigma$ with $\sigma=\pm$ are the occupancies of the corresponding spin-up/down states expressed through the products of creation, $d_{\sigma}^{\dag}$, and annihilation, $d_\sigma$, operators, $U$ is the on-cite Coulomb repulsion energy. $E_k$ describes the dispersion of electrons in the quantum wire with the wave vector $k$ along it, $n_{k,\sigma}=c_{k,\sigma}^\dag c_{k,\sigma}$ are the occupancies of the corresponding spin states in the wire, and $c_{k,\sigma}$ are the annihilation operators for these electrons. We assume the wire to be ballistic and neglect the interactions in it. The coefficients $V_{k,\sigma}$ describe the spin dependent tunneling between the quantum wire and the QD. Note, that the spin dependence in the form $V_{k,+}\neq V_{k,-}$ is allowed for any crystal structure of the host semiconductors, so the current induced spin accumulation is possible for a wide class of structures, including GaAs, Si, and Ge-based heterostructures. The time reversal symmetry imposes the relation $V_{k,+}=V_{-k,-}^*$.

This Hamiltonian is similar to the case of a QD coupled to the two leads, but here the two reservoirs are represented by the electrons moving from the left to the right and from the right to the left along the ballistic quantum wire~\cite{review1,teor1,teor2,teor3}. \addDima{This is mathematically equivalent to the standard configuration of a quantum dot coupled independently to the two leads with the different temperatures by the spin-dependent tunneling.} Consideration of only one size quantized electron level in the QD is the simplest approximation that allows us to describe the spin Nernst effect.

The Hamiltonian is diagonal in the spin space. Since tunneling of the electron with the opposite spins from the quantum wire to the QD leads to the opposite final spin states, this form of the Hamiltonian can be always achieved by appropriate choice of the spin basis at the QD. We neglect the spin flips inside the QD and the quantum wire assuming them to be slower than the tunneling rate. We also stress that we consider a nonmagnetic structure, where the spin dependence is produced by the spin-orbit coupling. If the crystal structure supports a vertical reflection plane that contains the quantum wire, then this reflection does not change direction of the heat current but flips the spins along vertical axis. As a result, if the QD is placed at the opposite side of the quantum wire, the heat current would produce the spin polarization in it with the opposite direction.  This behaviour qualitatively distinguish our system from the structures with magnetic leads, and for this reason the spin polarization induced by the heat flow can be called the tunneling spin Nernst effect.

\subsection{Formalism}



For the calculation of the current induced spin accumulation in the QD as a function of temperature gradient in the quantum wire and the Fermi level, we use the non-equilibrium Green's functions formalism~\cite{RevModPhys.58.323,stefanucci_vanleeuwen_2013,Arseev_2015}. The occupancies of the spin states in the QD are given by
\begin{equation}
  \label{eq:n}
  \braket{n_{\sigma}}=-\i\int\frac{\d\omega}{2\pi}G_{\sigma}^<(\omega),
\end{equation}
where $G_\sigma^<(\omega)$ are the lesser Green's functions of the QD (we set $\hbar=1$). They can be expressed through the corresponding lesser self energies $\Sigma_\sigma^<(\omega)$ and retarded Green's functions $G_\sigma^R(\omega)$ in the standard way:
\begin{equation}
  \label{eq:Gm}
  G_\sigma^<(\omega)=G_\sigma^R(\omega)\Sigma_\sigma^<(\omega)G_\sigma^{A}(\omega).
\end{equation}

To calculate it, we consider the bare Hubbard Green's function of a QD~\cite{doi:10.1098/rspa.1963.0204,haug2008quantum}
\begin{equation}
  \label{eq:G_R}
  G_{0,\sigma}^R(\omega)=\frac{1-\braket{n_{-\sigma}}}{\omega-E_0+\i\delta}+\frac{\braket{n_{-\sigma}}}{\omega-E_0-U+\i\delta}.
\end{equation}
To simplify the following we focus on the limit of strong Coulomb interaction, $U\to\infty$, and obtain~\cite{PhysRevLett.70.2601,Pols}
\begin{equation}
  \label{eq:G_R}
  G_{0,\sigma}^R(\omega)=\frac{1-\braket{n_{-\sigma}}}{\omega-E_0+\i\delta},
\end{equation}
where $\delta\rightarrow+0$. Then in the standard way using the wide flat band approximation~\cite{RevModPhys.58.323,stefanucci_vanleeuwen_2013,Arseev_2015}, we find from the Dyson equation the self energy $\Sigma_\sigma^R(\omega)=-\i\Gamma$~\cite{mahan2013many,Arseev_2015}, where $\Gamma=\Gamma_{L,\sigma}+\Gamma_{R,\sigma}$ is the spin independent total tunneling rate, which is related to the interaction with the electrons moving in the directions from the left and right leads, $\Gamma_{L,\sigma}$ and $\Gamma_{R,\sigma}$, respectively. They are related to the electrons with $k>0$ and $k<0$ and can be expressed as $\Gamma_{L/R,\sigma}=\pi D(E_0)\left|V_{\pm k_0,\sigma}^2\right|/4$, where $D(E_0)$ is the total density of states in the quantum wire at the QD energy and $k_0>0$ is the resonant wave vector defined by $E_{k_0}=E_0$. This gives the retarded Green's function
\begin{equation}
  \label{eq:G_R_main}
  G_{\sigma}^R(\omega)=\frac{1-\braket{n_{-\sigma}}}{\omega-E_0+\i(1-\braket{n_{-\sigma}})\Gamma}.
\end{equation}
The factor $1-\braket{n_{-\sigma}}$ here describes the suppression of tunneling of electrons with the opposite spin to the QD by the Coulomb interaction.

In the same approximation we find the lesser self energy (we remind that we denote with $\Gamma$'s amplitude decay rates)
\begin{equation}
  \label{eq:sigma}
  \Sigma_\sigma^<(\omega)=2\i\left[\Gamma_{L,\sigma} f_L(\omega)+\Gamma_{R,\sigma}f_R(\omega)\right],
\end{equation}
where the distribution function of the electron moving from the left and right leads are
\begin{equation}
  \label{eq:fLR}
  f_{L/R}(E)=\frac{1}{1+\exp\left[(E-E_F)/T_{L/R}\right]}
\end{equation}
with $E_F$ being the Fermi energy and $T_{L/R}$ being the temperatures of the left/right leads ($k_B=1$). Substituting Eqs.~\eqref{eq:G_R_main} and~\eqref{eq:sigma} in Eqs.~\eqref{eq:n} and~\eqref{eq:Gm} we obtain a closed set of two integral equations which determine occupation numbers $\langle n_{\sigma}\rangle$ for spin-up and spin-down states in the QD.

This corresponds to the Hartree-Fock approximation~\cite{Lacroix_1981,Mantsevich2022,Arseyev2012,Mantsevich2013,Mantsevich2010}, which neglects correlations between electrons in the quantum wire, or to the method of equations of motion~\cite{PhysRevLett.70.2601,haug2008quantum,Arseev2012,Arseyev2011}. Thus the applicability of this approach is limited to the temperatures above the Kondo temperature~\cite{PhysRevLett.70.2601,Niu_1999,Mantsevich2023}. Also small size of QDs is required, so that only the lowest level of size quantization can be considered, and temperature should be smaller than the energy spacings between the levels in order to neglect thermal population of the excited states in the system.

The spin-orbit interactions gives rise to a small difference of the tunneling matrix elements and corresponding difference of the tunneling rates, which we present as $\Gamma_{L,+}=\Gamma_{R,-}=\Gamma/2+\gamma$ and $\Gamma_{R,+}=\Gamma_{L,-}=\Gamma/2-\gamma$ with $\gamma\ll\Gamma$ being a phenomenological spin-dependent contribution. Then the equations for the occupancies read
\begin{widetext}
  \begin{equation}
    \label{eq:Coulomb_1}
    \langle n_{\sigma}\rangle=\frac{1}{\pi}\int\frac{(1-\langle n_{-\sigma}\rangle)^{2}}{(\omega-E_0)^2+\Gamma^2(1-\langle n_{-\sigma}\rangle)^{2}}
    \left[\left(\frac{\Gamma}{2}+\sigma\gamma\right)\frac{1}{1+e^{\frac{\omega-E_F}{T_L}}}+\left(\frac{\Gamma}{2}-\sigma\gamma\right)\frac{1}{1+e^{\frac{\omega-E_F}{T_R}}}\right]
    d\omega.
  \end{equation}
\end{widetext}
We are looking for the bilinear response of the spin polarization in the QD $S=\left(\langle n_+\rangle-\langle n_-\rangle\right)/2$ to $\gamma$ and the temperature gradient. So we consider $\langle n_{\sigma}\rangle=\braket{n}+\sigma S$ and $T_{L/R}=T\pm\delta T/2$, where $\braket{n}$ and $T$ are average occupancy and temperature, respectively, and $\delta T$ is the difference of the temperatures in the left and right leads. The average occupancy $\braket{n}$ can be found neglecting spin-orbit contribution $\gamma$, temperature difference $\delta T$, and spin polarization $S$ in Eq.~\eqref{eq:Coulomb_1}:
\begin{equation}
  \label{eq:n_sol}
\langle n\rangle=\frac{\Gamma}{\pi}(1-\langle n\rangle)^{2}\int\frac{\left[1+e^{\frac{\omega-E_F}{T}}\right]^{-1}}{(\omega-E_0)^2+\Gamma^2(1-\langle n\rangle)^{2}}\d\omega.
\end{equation}
Then we calculate the accumulated spin as $S=\left(\braket{n_+}-\braket{n_-}\right)/2$ using Eq.~\eqref{eq:Coulomb_1}. In the right-hand side we substitute the above expressions for $\braket{n_\sigma}$ and $T_{L,R}$, and then keep only the contributions linear in $S$ or in $\delta T$. This gives
\begin{multline}
  S={\frac{2S}{\pi}\int\frac{\Gamma(1-\langle n\rangle)(\omega-E_0)^{2}\d\omega}{[(\omega-E_0)^{2}+\Gamma^{2}(1-\langle n\rangle)^{2}]^{2}\left(1+e^{\frac{\omega-E_F}{T}}\right)}}
  \\+
  {\frac{\gamma\delta T}{4\pi T^{2}}\int\frac{(1-\langle n\rangle)^{2}(\omega-E_F)\d\omega}{[(\omega-E_0)^{2}+\Gamma^{2}(1-\langle n\rangle)^{2}]\cosh^{2}(\frac{\omega-E_F}{2T})}}.
\end{multline}
So the electron spin can be calculated as
\begin{widetext}
\begin{equation}
  \label{eq:n_sol_1}
  S={\frac{\gamma\delta T}{4\pi T^{2}}\int\frac{(1-\langle n\rangle)^{2}(\omega-E_F)\d\omega}{[(\omega-E_0)^{2}+\Gamma^{2}(1-\langle n\rangle)^{2}]\cosh^{2}(\frac{\omega-E_F}{2T})}}
  \left/
  \left\{
    {1-\frac{2}{\pi}\int\frac{\Gamma(1-\langle n\rangle)(\omega-E_0)^{2}\d\omega}{[(\omega-E_0)^{2}+\Gamma^{2}(1-\langle n\rangle)^{2}]^{2}\left(1+e^{\frac{\omega-E_F}{T}}\right)}}
  \right\}
  \right.
  .
\end{equation}
\end{widetext}
The denominator here describes the enhancement of spin polarization in the QD by the Coulomb interaction. Physically, this happens because a spin-up electron in the QD forbids tunneling of spin-down electron to it for the large Coulomb interaction. Also, one can see that the spin polarization vanishes at $E_F=E_0$ because of the symmetry of the Fermi distribution functions in this case. Generally, the integrals here can not be solved analytically.

It is useful to consider expression for the electric current along the quantum wire
  \begin{equation}
    I=\frac{e}{\pi}\int\left[f_L(\omega)-f_R(\omega)\right]\d\omega,
  \end{equation}
  where $e$ is the electron charge. Substituting here the distributions functions~\eqref{eq:fLR}, we obtain in the first order in $\delta T$
  \begin{equation}
    \label{eq:Itot}
    I=\frac{e\delta T}{4\pi T^2}\int\frac{\omega-E_F}{\cosh^2\left(\frac{\omega-E_F}{2T}\right)}\d\omega.
  \end{equation}
  Now from the comparison with Eq.~\eqref{eq:n_sol_1} one can see that the spin polarization induced by the heat flow is qualitatively proportional to the electric current at the energy of the QD, $E_0$. In particular, it vanishes at $E_0=E_F$. We note also that in Eq.~\eqref{eq:Itot} there is no total electric current, as required for the spin Nernst effect.

\subsection{Illustrative example}

We stress that the heat induced spin polarization appears due to the spin-orbit interaction, which does not require breaking of the time reversal symmetry. As a result, the spin polarization is an odd function of the temperature bias, and also of the distance between the QD and the quantum wire, if there is a vertical mirror reflection along the quantum wire.

This can be illustrated considering a specific Hamiltonian of a quantum wire with Rashba spin-orbit interaction:
  \begin{equation}
    \label{eq:HQW}
    \mathcal H_{QW}=\frac{\bm k^2}{2m}-U_2\delta(y)+\alpha(\sigma_x k_y-\sigma_y k_x).
  \end{equation}
  Here $\bm k=-\i\bm\nabla$ is the electron spin momentum operator, $m$ is its effective mass, $\alpha$ is the spin-orbit interaction constant, $\bm\sigma$ are the Pauli matrices, and $U_2$ is the strength of attractive delta-potential of the quantum wire. The eigenstates of this Hamiltonian are found in Appendix~\ref{sec:spin_polarization}. For the given energy $E$ there are four degenerate states corresponding to the electrons moving to the right and to the left with the spin along the $y$ axis and in the opposite direction.

\begin{figure}
\includegraphics[width=0.8\linewidth]{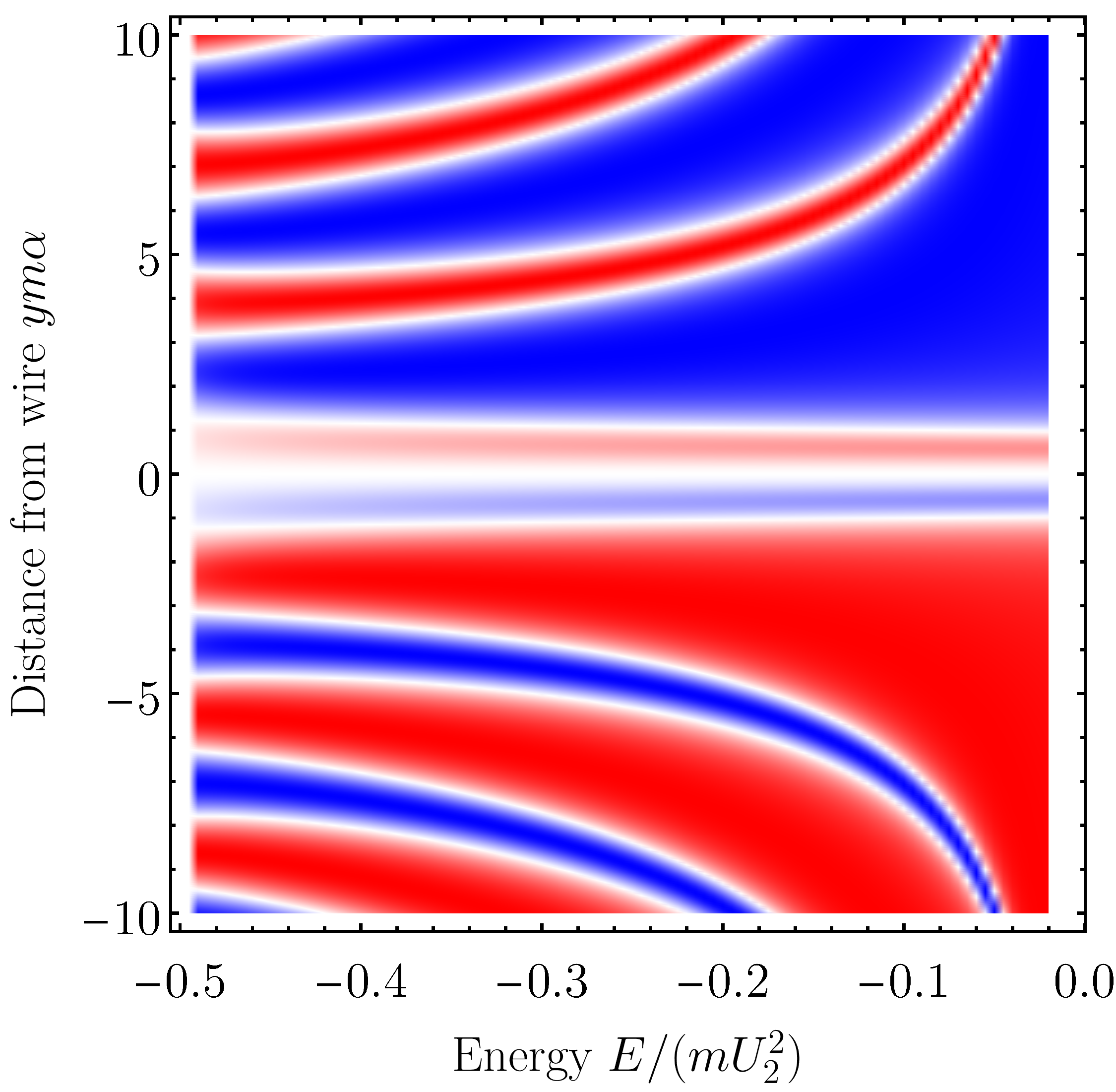}
\caption{
  \label{fig:polarization}
  Average spin polarization of electrons tunneling with the energy $E$ and $k_x>0$ to the distance $y$ from the quantum wire calculated for the Hamiltonian~\eqref{eq:HQW} with $\alpha/U_2=0.1$. Blue and red colors correspond to $P_z=\pm1$.
}
\end{figure}

Let us consider the given energy and the two states of electron moving to the right $\Psi_\uparrow(x,y)$ and $\Psi_\downarrow(x,y)$. Then we calculate the average spin polarization along $z$ direction as a function of distance from the quantum wire:
  \begin{equation}
    \label{eq:Pz}
    P_z(y)=\frac{\left<\Psi_\uparrow\middle|\sigma_z\middle|\Psi_\uparrow\right>+\left<\Psi_\downarrow\middle|\sigma_z\middle|\Psi_\downarrow\right>}{\left<\Psi_\uparrow\middle|\Psi_\uparrow\right>+\left<\Psi_\downarrow\middle|\Psi_\downarrow\right>}.
\end{equation}
It does not depend on $x$ since the wave functions are the plain waves along this direction. The spin polarization is shown in Fig.~\ref{fig:polarization}. It is an odd function of $y$ in agreement with the symmetry of the Hamiltonian~\eqref{eq:HQW}. This underlines that the spin polarization appears due to the spin Nernst effect and is not related to the spin injection from the contacts. For small distances and weak spin-orbit coupling, the spin polarization is proportional to $\alpha^2$, but at large distances from the quantum wire it can reach $\pm1$.

We note that other example of calculations of the spin dependent tunneling rates can be found, for example, in Refs.~\onlinecite{tarasenko:056601,Hopping_spin,Mantsevich2022}.

\begin{figure*}
\includegraphics[width=\linewidth]{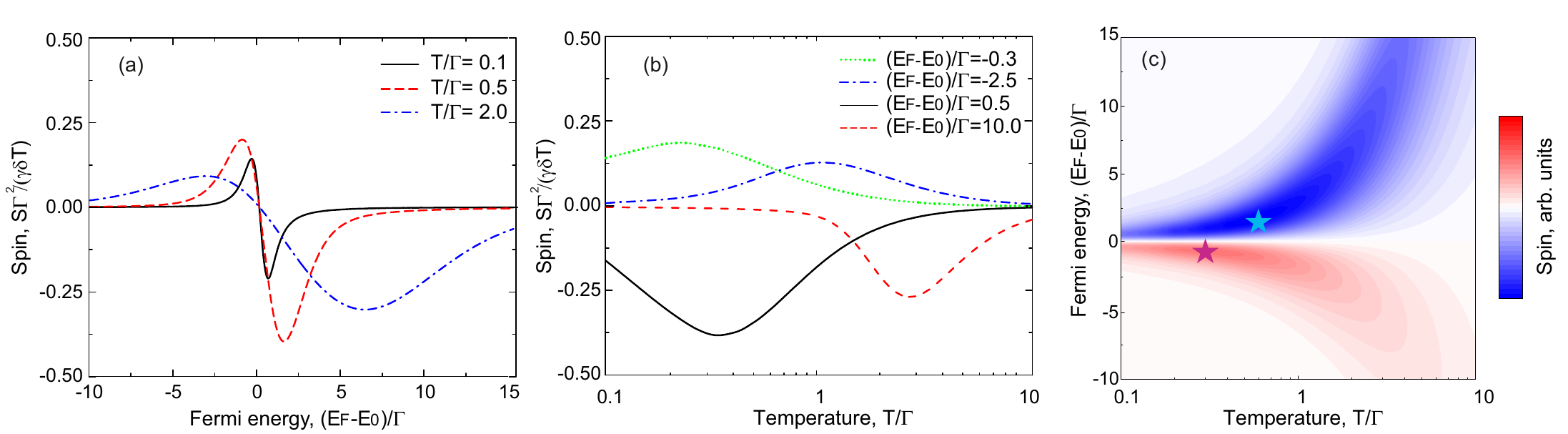}
\caption{
  \label{fig:Spin_3D_Coulomb_1}
  Electron spin in the quantum dot induced by the heat flow along the quantum wire calculated after Eqs.~\eqref{eq:n_sol} and~\eqref{eq:n_sol_1} as a function of Fermi energy for the temperatures given in the legend (a), as a function of temperature for the Fermi energies given in the legend (b), and as a function of both parameters (c). The cyan and red magenta stars show the minimum and maximum, respectively.
}
\end{figure*}

\section{Results and Discussion} 
\label{sec:results}

The electron spin in the quantum dot induced by the heat flow along the quantum wire is shown in Fig.~\ref{fig:Spin_3D_Coulomb_1}. The calculations are performed by numerical solution of Eqs.~\eqref{eq:n_sol} and~\eqref{eq:n_sol_1}, which describe the limit of strong Coulomb interaction. Panel (a) shows that the spin vanishes for zero, large positive and large negative Fermi energies and changes sign at $E_F=E_0$. Generally, it reaches the largest absolute values at $|E_F-E_0|\sim\Gamma$ with the coefficient increasing with the growth of temperature.

The temperature dependence is shown in more detail in Fig.~\ref{fig:Spin_3D_Coulomb_1}(b) for a few different Fermi energies. One can see that the spin vanishes in the limits of small and large temperatures, and reaches the larges absolute value at $T\sim\Gamma$. The overall dependence of spin on the Fermi energy and the temperature is shown in the color map~\ref{fig:Spin_3D_Coulomb_1}(c). Here the magenta star show the maximum (at $\gamma\delta T>0$) spin $S=0.212\gamma\delta T/\Gamma^2$, which is reached at $E_F-E_0=-0.6\Gamma$ and $T=0.3\Gamma$. Similarly, the cyan star shows the minimum of spin $S=-0.397\gamma\delta T/\Gamma^2$, which is reached at $E_F-E_0=1.5\Gamma$ and $T=0.6\Gamma$. Thus, the largest spin is generally reached at $|E_F-E_0|,T\sim\Gamma$.

This can be also seen from analytical expressions for a few limiting cases. For example, at the low temperatures $T\ll\Gamma$ and large Fermi energies $E_F-E_0\gg\Gamma$, the occupancy of the QD equals one half, $\langle n\rangle=1/2$, so Eq.~\eqref{eq:n_sol_1} gives
\begin{equation}
  \label{eq:lim1}
S=-\frac{\pi^{2}\gamma T\delta T}{6\Gamma(E_F-E_0)^{2}}.
\end{equation}
In the opposite limit of low Fermi energy, $E_0-E_F\gg\Gamma$, the QD occupancy is very low $\langle n\rangle\ll1$, so we obtain
\begin{equation}
  \label{eq:lim2}
S=\frac{2\pi\gamma T\delta T}{3(E_0-E_F)^{3}}.
\end{equation}
From these two limits one can see that the spin has opposite signs for $E_F>E_0$ and $E_F<E_0$, and that it vanishes for $|E_F-E_0|\gg\Gamma$. The spin polarization in Eqs.~\eqref{eq:lim1} and~\eqref{eq:lim2} linearly increases with increase of the temperature. The limit $|E_0-E_F|\gg T\gg\Gamma$ is described by these expressions as well.

In the limit of high temperatures $T\gg\Gamma,|E_F-E_0|$, Eq.~\eqref{eq:n_sol} gives occupancy of the QD $\langle n\rangle=1/3$. Using it, we obtain from Eq.~\eqref{eq:n_sol_1} the spin
\begin{equation}
S=\frac{\gamma(E_0-E_F)\delta T}{3\Gamma T^{2}}.
\end{equation}
It again changes sign at $E_F=E_0$, but also shows a decrease at large temperatures $\propto 1/T^2$. So the maximum and minimum at $|E_F-E_0|,T\sim\Gamma$ can be seen indeed.

\begin{figure}
\includegraphics[width=0.75\linewidth]{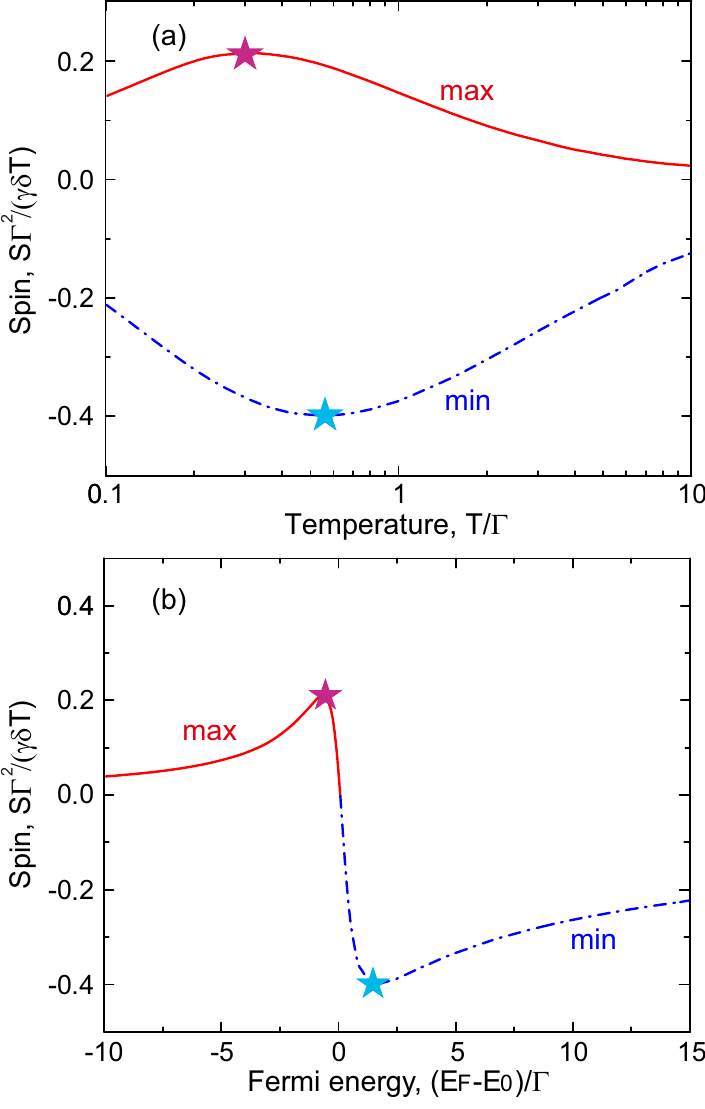}
\caption{\label{fig:Spin_max}
  (a) Maximum (red solid curve) and minimum (blue dash-dotted curve) of spin as a function of Fermi energy for the given temperature. (b) Maximum (red solid curve) and minimum (blue dash-dotted curve) of spin as a function of temperature for the given Fermi energy.}
\end{figure}

We also find it useful to show the maximum and minimum of spin as functions of Fermi energy for a given temperature in Fig.~\ref{fig:Spin_max}(a). Similarly, the extrema of spin as functions of temperature for the given Fermi energy are shown in Fig.~\ref{fig:Spin_max}(b). The cyan and magenta stars in this figure show the same extrema as in Fig.~\ref{fig:Spin_3D_Coulomb_1}(c).

Generally, the spin in the QD induced by the heat flow is of the order of $\gamma\delta T/\Gamma^2$, so in the framework of our theory it is always small. However, in realistic structures with $\Gamma\sim 10~\mu$eV, the temperatures difference in the left and right leads can be of the same order for $\delta T\sim0.1$~K. So the linear response theory may be insufficient, and one can expect that for the large $\delta T$ spin will be of the order of $\gamma/\Gamma$. Moreover, Fig.~\ref{fig:polarization} shows that at significant distances from the quantum wire, spin dependence of the tunneling rate can be very strong. This again violates our assumption of $\gamma/\Gamma\ll 1$, but we speculate that the heat flow induced spin accumulation in this case would be large. In principle, it can be of the order of $100\%$ as there is no small parameter in the system anymore.

Note that the typical tunneling rates are much faster than the spin relaxation of localized electrons, so the latter does not play a role. The spin polarization can be then measured optically using polarized luminescence or spin-induced Faraday rotation and electrically using magnetic point contacts.

\section{Conclusion}
\label{sec:conclusion}

We have shown that the spin-orbit interaction produces the spin Nernst effect in a gate defined heterostructure consisting of a QD side coupled to a quantum wire without magnetic elements. The heat current produces spin polarization of a localized electron in this system due to the spin dependent tunneling. The spin polarization is enhanced by the Coulomb interaction in the QD. It is a nonmonotonous function of Fermi energy and temperature, and reaches the largest values when both are of the order of the tunneling rate between the QD and the quantum wire. The spin accumulation in the framework of our model is parametrically weak, but continuation of the theory to large temperature bias allows one to expect macroscopically large spin polarization.

\section{Acknowledgements}

We would like to thank S.A. Tarasenko and D.A. Frolov for useful discussions. We thank the Foundation for the Advancement of Theoretical Physics and Mathematics ``BASIS''. Analytical calculations by D.S.S. were supported by the Russian Science Foundation grant No. 23-12-00142. Numerical modeling by V.N.M. were supported by the Russian Science Foundation grant No. 22-12-00139.

\appendix

\section{Eigenfunction of Eq.~\eqref{eq:HQW}}
\label{sec:spin_polarization}

The Hamiltonian~\eqref{eq:HQW} is translationally invariant along the $x$ axis, so its eigenfunctions are proportional to $\e^{\i kx}$. Moreover, due to the assumption of delta-potential of the quantum wire, the wave function away from it (at $y\neq0$) can be written as
\begin{equation}
  \Psi=\begin{pmatrix}
\chi_{\uparrow}\\
\chi_{\downarrow}
\end{pmatrix}
\e^{\i kx-\kappa|y|}.
\end{equation}
Substituting this form to Eq.~\eqref{eq:HQW} we obtain
\begin{eqnarray}
E\chi_{\uparrow}=\frac{k^2-\kappa^{2}}{2m}\chi_{\uparrow}+\i\alpha k\chi_{\downarrow}+\i\alpha\kappa\chi_{\downarrow},\nonumber\\
E\chi_{\downarrow}=\frac{k^2-\kappa^{2}}{2m}\chi_{\downarrow}-\i\alpha k_{x}\chi_{\uparrow}+\i\alpha\kappa\chi_{\uparrow}.
\label{eq:system_1}
\end{eqnarray}
for $y>0$, while for $y<0$ the sign of $\kappa$ should be reversed. Here $E<0$ is the electron energy. These equations have two solutions with the two inverse decay lengths $\kappa_\pm$, which are determined by
\begin{equation}
  \kappa_\pm^2-\varkappa^2-k^2=\mp2\i\beta\sqrt{\kappa_\pm^2-k^2},
\end{equation}
with $\varkappa=\sqrt{-2Em}$ and $\beta=\alpha m$. Also from Eq.~\eqref{eq:system_1} we find the corresponding ratios of the spinor components
\begin{subequations}
  \begin{equation}
    \chi_\uparrow/\chi_\downarrow=\mp\xi_\pm,
  \end{equation}
  \begin{equation}
    \chi_\downarrow/\chi_\uparrow=\pm\xi_\pm,
  \end{equation}
\end{subequations}
at $y>0$ and $y<0$, respectively, with $\xi_\pm=\sqrt{(\kappa_\pm+k)/(\kappa_\pm-k)}$.

The eigenfunctions have the form of linear combinations of the these solutions:
\begin{multline}
  \label{eq:wf}
  \Psi=\sum_\pm \left[A_\pm
  \begin{pmatrix}
\mp\xi_\pm\\
1
\end{pmatrix}
\e^{-\kappa_\pm y}\theta(y)
\right.\\+\left.
B_\pm
\begin{pmatrix}
1\\
\pm\xi_\pm
\end{pmatrix}
\e^{\kappa_\pm y}\theta(-y)
\right]\e^{\i kx}.
\end{multline}
The coefficients $A_\pm$ and $B_\pm$ can be found from the boundary conditions at $y=0$. Firstly, the wave function should be continuous, which can be written as
\begin{subequations}
  \begin{equation}
    \left.\Psi\right|^{+0}_{-0}=0.
  \end{equation}
  And integrating Schrodinger equation around $y=0$ we obtain the boundary condition for the wave function derivative:
  \begin{equation} 
    \left.\frac{\d\Psi}{\d y}\right|^{+0}_{-0}=-2mU_2\Psi
  \end{equation}
  (the wave function should be taken at $y=0$ and it is continuous at it).
\end{subequations}
Substitution of the wave function~\eqref{eq:wf} to the boundary conditions yields four linear equations for the coefficients $A_\pm$ and $B_\pm$:
\begin{subequations}
  \label{eq:system}
    \begin{equation}
      -\xi_{+}A_{+}+\xi_{-}A_{-}=B_{+}+B_{-},
    \end{equation}
    \begin{equation}
      A_{+}+A_{-}=\xi_{+}B_{+}-\xi_{-}B_{-},
    \end{equation}
    \begin{equation}
      -\kappa_+\xi_+A_++\kappa_-\xi_-A_-+\kappa_+B_++\kappa_-B_-=2p(B_++B_-),
    \end{equation}
    \begin{equation}
     \kappa_+A_++\kappa_-A_-+\kappa_+\xi_+B_+-\kappa_-\xi_-B_-=2p(A_++A_-),
    \end{equation}
  \end{subequations}
  where $p=mU_2$. The determinant of this system should equal zero
\begin{multline}
  [-\xi_+^{2}(\kappa_++\kappa_-)-2\xi_+\xi_-\kappa_++\kappa_+-\kappa_-+2p\xi_+^{2}+2p\xi_+\xi_-]\\
  \times[\xi_+\kappa_-+\xi_-\kappa_-+\xi_-^{2}\xi_+\kappa_++\xi_+\kappa_+-\xi_-^{2}\xi_+\kappa_-+\xi_-\kappa_--2p\xi_+\\-2p\xi_-]
  -[\xi_-^{2}(\kappa_++\kappa_-)+2\xi_+\xi_-\kappa_-+\kappa_+-\kappa_--2p\xi_-^{2}\\-2p\xi_+\xi_-]
  \times[\xi_+\kappa_++\xi_-\kappa_+-\xi_+^{2}\xi_-\kappa_++\xi_+\kappa_++\xi_+^{2}\xi_-\kappa_-\\+\xi_-\kappa_--2p\xi_+-2p\xi_-]
  =0,
\label{eq:determinant}
\end{multline}
which gives the relation between the wave vector $k$ and energy $E$. Generally, there are two solutions of this equation, and they give the two wave functions $\Psi_\uparrow$ and $\Psi_\downarrow$.

For example, for $k=0$ we obtain $\xi_{+}=\xi_{-}=1$ and 
\begin{eqnarray}
\kappa_{\pm}=p\mp\i\beta.
\end{eqnarray}
The determinant~\eqref{eq:determinant} equals zero when $\kappa_{+}+\kappa_{-}=2p$. This condition directly determines the energy of the ground state $E=-m(U_{2}^{2}+\alpha^2)/2$. It is twofold degenerate, so there are two linearly independent solutions of Eqs.~\eqref{eq:system}. Taking into account the normalization, they can be chosen in the form
\begin{eqnarray}
&&A_{+}=-A_{-}=-\sqrt{\frac{p}{4L}},\nonumber\\
&&B_{+}=B_{-}=\sqrt{\frac{p}{4L}},
\end{eqnarray}
and
\begin{eqnarray}
&&A_{+}=A_{-}=\sqrt{\frac{p}{4L}},\nonumber\\
&&B_{+}=-B_{-}=\sqrt{\frac{p}{4L}}.
\end{eqnarray}
where $L$ is the normalization length along the quantum wire. Substituting these coefficients to Eqs.~\eqref{eq:wf}, we get an explicit expressions for spin up and spin down wave functions:
\begin{subequations}
  \begin{equation}
    \Psi_{\uparrow}=\sqrt{\frac{p}{L}}\e^{-p|y|}
      \begin{pmatrix}
        \cos(\beta y)\\
        -\i\sin(\beta y)
      \end{pmatrix},
  \end{equation}
  \begin{equation}
    \Psi_{\downarrow}=\sqrt{\frac{p}{L}}\e^{-p|y|}
      \begin{pmatrix}
        -\i\sin(\beta y)\\
        \cos(\beta y)
      \end{pmatrix}.
  \end{equation}
\end{subequations}
One can see that the spin polarization in these states depends on the distance from the quantum wire as $S_z\propto\pm\cos(2\beta y)$, $S_y\propto\mp\sin(2\beta y)$, $S_x=0$. The spin precession is directly related to the spin-orbit coupling in Eq.~\eqref{eq:HQW}. At the same time, there is no spin polarization on average in the states with $k=0$.

Generally, in the first order in the spin-orbit coupling we obtain from Eq.~\eqref{eq:determinant} for the given energy the two positive wave vectors $k=\sqrt{p^2-\varkappa^2}+\beta$ and $\sqrt{p^2-\varkappa^2}-\beta$. For these states we obtain
\begin{subequations}
  \begin{equation}
    \kappa_\pm=p\mp\i\frac{\varkappa}{p}\beta+\sqrt{1-\frac{\varkappa^2}{p^2}}\beta,
  \end{equation}
  \begin{equation}
    \kappa_\pm=p\mp\i\frac{\varkappa}{p}\beta-\sqrt{1-\frac{\varkappa^2}{p^2}}\beta,
  \end{equation}
\end{subequations}
respectively. One can see from the last term that the two contributions to the wave function decay at slightly different lengths. This difference appears in the first order in $\beta$, and the spin rotation is also proportional to $\beta$, so the spin polarization [Eq.~\eqref{eq:Pz}] appears in the second order in the spin-orbit interaction.

Note that the definition of spin polarization in Eq.~\eqref{eq:Pz} assumes equal occupancies of the states $\Psi_\uparrow$ and $\Psi_\downarrow$ due to the coincidence of their energies. This results in a weak current of spin along $y$ axis in the direction of the quantum wire. But this effect appears in the higher orders in $\beta$ and its exclusion does not change, for example, Fig.~\ref{fig:polarization}.

\renewcommand{\i}{\ifr}


\begin{thebibliography}{37}%
\makeatletter
\providecommand \@ifxundefined [1]{%
 \@ifx{#1\undefined}
}%
\providecommand \@ifnum [1]{%
 \ifnum #1\expandafter \@firstoftwo
 \else \expandafter \@secondoftwo
 \fi
}%
\providecommand \@ifx [1]{%
 \ifx #1\expandafter \@firstoftwo
 \else \expandafter \@secondoftwo
 \fi
}%
\providecommand \natexlab [1]{#1}%
\providecommand \enquote  [1]{``#1''}%
\providecommand \bibnamefont  [1]{#1}%
\providecommand \bibfnamefont [1]{#1}%
\providecommand \citenamefont [1]{#1}%
\providecommand \href@noop [0]{\@secondoftwo}%
\providecommand \href [0]{\begingroup \@sanitize@url \@href}%
\providecommand \@href[1]{\@@startlink{#1}\@@href}%
\providecommand \@@href[1]{\endgroup#1\@@endlink}%
\providecommand \@sanitize@url [0]{\catcode `\\12\catcode `\$12\catcode
  `\&12\catcode `\#12\catcode `\^12\catcode `\_12\catcode `\%12\relax}%
\providecommand \@@startlink[1]{}%
\providecommand \@@endlink[0]{}%
\providecommand \url  [0]{\begingroup\@sanitize@url \@url }%
\providecommand \@url [1]{\endgroup\@href {#1}{\urlprefix }}%
\providecommand \urlprefix  [0]{URL }%
\providecommand \Eprint [0]{\href }%
\providecommand \doibase [0]{https://doi.org/}%
\providecommand \selectlanguage [0]{\@gobble}%
\providecommand \bibinfo  [0]{\@secondoftwo}%
\providecommand \bibfield  [0]{\@secondoftwo}%
\providecommand \translation [1]{[#1]}%
\providecommand \BibitemOpen [0]{}%
\providecommand \bibitemStop [0]{}%
\providecommand \bibitemNoStop [0]{.\EOS\space}%
\providecommand \EOS [0]{\spacefactor3000\relax}%
\providecommand \BibitemShut  [1]{\csname bibitem#1\endcsname}%
\let\auto@bib@innerbib\@empty
\bibitem [{\citenamefont {Bose}\ and\ \citenamefont
  {Tulapurkar}(2019)}]{bose19}%
  \BibitemOpen
  \bibfield  {author} {\bibinfo {author} {\bibfnamefont {A.}~\bibnamefont
  {Bose}}\ and\ \bibinfo {author} {\bibfnamefont {A.~A.}\ \bibnamefont
  {Tulapurkar}},\ }\bibfield  {title} {\bibinfo {title} {{Recent advances in
  the spin Nernst effect}},\ }\href
  {https://doi.org/10.1016/j.jmmm.2019.165526} {\bibfield  {journal} {\bibinfo
  {journal} {J. Magn. Magn. Mater.}\ }\textbf {\bibinfo {volume} {491}},\
  \bibinfo {pages} {165526} (\bibinfo {year} {2019})}\BibitemShut {NoStop}%
\bibitem [{\citenamefont {Cheng}\ \emph {et~al.}(2008)\citenamefont {Cheng},
  \citenamefont {Xing}, \citenamefont {Sun},\ and\ \citenamefont
  {Xie}}]{cheng08}%
  \BibitemOpen
  \bibfield  {author} {\bibinfo {author} {\bibfnamefont {S.-g.}\ \bibnamefont
  {Cheng}}, \bibinfo {author} {\bibfnamefont {Y.}~\bibnamefont {Xing}},
  \bibinfo {author} {\bibfnamefont {Q.-f.}\ \bibnamefont {Sun}},\ and\ \bibinfo
  {author} {\bibfnamefont {X.~C.}\ \bibnamefont {Xie}},\ }\bibfield  {title}
  {\bibinfo {title} {{Spin Nernst effect and Nernst effect in two-dimensional
  electron systems}},\ }\href {https://doi.org/10.1103/PhysRevB.78.045302}
  {\bibfield  {journal} {\bibinfo  {journal} {Phys. Rev. B}\ }\textbf {\bibinfo
  {volume} {78}},\ \bibinfo {pages} {045302} (\bibinfo {year}
  {2008})}\BibitemShut {NoStop}%
\bibitem [{\citenamefont {Liu}\ and\ \citenamefont {Xie}(2010)}]{liu10}%
  \BibitemOpen
  \bibfield  {author} {\bibinfo {author} {\bibfnamefont {X.}~\bibnamefont
  {Liu}}\ and\ \bibinfo {author} {\bibfnamefont {X.~C.}\ \bibnamefont {Xie}},\
  }\bibfield  {title} {\bibinfo {title} {{Spin Nernst effect in the absence of
  a magnetic fields}},\ }\href {https://doi.org/10.1016/j.ssc.2009.12.017}
  {\bibfield  {journal} {\bibinfo  {journal} {Solid State Commum.}\ }\textbf
  {\bibinfo {volume} {150}},\ \bibinfo {pages} {471} (\bibinfo {year}
  {2010})}\BibitemShut {NoStop}%
\bibitem [{\citenamefont {Dyakonov}\ and\ \citenamefont
  {Perel'}(1971)}]{dyakonov71}%
  \BibitemOpen
  \bibfield  {author} {\bibinfo {author} {\bibfnamefont {M.~I.}\ \bibnamefont
  {Dyakonov}}\ and\ \bibinfo {author} {\bibfnamefont {V.~I.}\ \bibnamefont
  {Perel'}},\ }\bibfield  {title} {\bibinfo {title} {Possibility of orienting
  electron spins with current},\ }\href@noop {} {\bibfield  {journal} {\bibinfo
   {journal} {JETP Lett.}\ }\textbf {\bibinfo {volume} {13}},\ \bibinfo {pages}
  {467} (\bibinfo {year} {1971})}\BibitemShut {NoStop}%
\bibitem [{\citenamefont {Tauber}\ \emph {et~al.}(2012)\citenamefont {Tauber},
  \citenamefont {Gradhand}, \citenamefont {Fedorov},\ and\ \citenamefont
  {Mertig}}]{tauber12}%
  \BibitemOpen
  \bibfield  {author} {\bibinfo {author} {\bibfnamefont {K.}~\bibnamefont
  {Tauber}}, \bibinfo {author} {\bibfnamefont {M.}~\bibnamefont {Gradhand}},
  \bibinfo {author} {\bibfnamefont {D.~V.}\ \bibnamefont {Fedorov}},\ and\
  \bibinfo {author} {\bibfnamefont {I.}~\bibnamefont {Mertig}},\ }\bibfield
  {title} {\bibinfo {title} {{Extrinsic Spin Nernst Effect from First
  Principles}},\ }\href {https://doi.org/10.1103/PhysRevLett.109.026601}
  {\bibfield  {journal} {\bibinfo  {journal} {Phys. Rev. Lett.}\ }\textbf
  {\bibinfo {volume} {109}},\ \bibinfo {pages} {026601} (\bibinfo {year}
  {2012})}\BibitemShut {NoStop}%
\bibitem [{\citenamefont {Wimmer}\ \emph {et~al.}(2013)\citenamefont {Wimmer},
  \citenamefont {K\"odderitzsch}, \citenamefont {Chadova},\ and\ \citenamefont
  {Ebert}}]{wimmer13}%
  \BibitemOpen
  \bibfield  {author} {\bibinfo {author} {\bibfnamefont {S.}~\bibnamefont
  {Wimmer}}, \bibinfo {author} {\bibfnamefont {D.}~\bibnamefont
  {K\"odderitzsch}}, \bibinfo {author} {\bibfnamefont {K.}~\bibnamefont
  {Chadova}},\ and\ \bibinfo {author} {\bibfnamefont {H.}~\bibnamefont
  {Ebert}},\ }\bibfield  {title} {\bibinfo {title} {{First-principles linear
  response description of the spin Nernst effect}},\ }\href
  {https://doi.org/10.1103/PhysRevB.88.201108} {\bibfield  {journal} {\bibinfo
  {journal} {Phys. Rev. B}\ }\textbf {\bibinfo {volume} {88}},\ \bibinfo
  {pages} {201108} (\bibinfo {year} {2013})}\BibitemShut {NoStop}%
\bibitem [{\citenamefont {Guo}\ and\ \citenamefont {Wang}(2017)}]{guo17}%
  \BibitemOpen
  \bibfield  {author} {\bibinfo {author} {\bibfnamefont {G.-Y.}\ \bibnamefont
  {Guo}}\ and\ \bibinfo {author} {\bibfnamefont {T.-C.}\ \bibnamefont {Wang}},\
  }\bibfield  {title} {\bibinfo {title} {{Large anomalous Nernst and spin
  Nernst effects in the noncollinear antiferromagnets ${\mathrm{Mn}}_{3}X$
  ($X=\mathrm{Sn},\mathrm{Ge},\mathrm{Ga}$)}},\ }\href
  {https://doi.org/10.1103/PhysRevB.96.224415} {\bibfield  {journal} {\bibinfo
  {journal} {Phys. Rev. B}\ }\textbf {\bibinfo {volume} {96}},\ \bibinfo
  {pages} {224415} (\bibinfo {year} {2017})}\BibitemShut {NoStop}%
\bibitem [{\citenamefont {Kovalev}\ and\ \citenamefont
  {Zyuzin}(2016)}]{kovalev16}%
  \BibitemOpen
  \bibfield  {author} {\bibinfo {author} {\bibfnamefont {A.~A.}\ \bibnamefont
  {Kovalev}}\ and\ \bibinfo {author} {\bibfnamefont {V.}~\bibnamefont
  {Zyuzin}},\ }\bibfield  {title} {\bibinfo {title} {{Spin torque and Nernst
  effects in Dzyaloshinskii-Moriya ferromagnets}},\ }\href
  {https://doi.org/10.1103/PhysRevB.93.161106} {\bibfield  {journal} {\bibinfo
  {journal} {Phys. Rev. B}\ }\textbf {\bibinfo {volume} {93}},\ \bibinfo
  {pages} {161106} (\bibinfo {year} {2016})}\BibitemShut {NoStop}%
\bibitem [{\citenamefont {Shitade}(2022)}]{PhysRevB.106.045203}%
  \BibitemOpen
  \bibfield  {author} {\bibinfo {author} {\bibfnamefont {A.}~\bibnamefont
  {Shitade}},\ }\bibfield  {title} {\bibinfo {title} {{Spin accumulation in the
  spin Nernst effect}},\ }\href {https://doi.org/10.1103/PhysRevB.106.045203}
  {\bibfield  {journal} {\bibinfo  {journal} {Phys. Rev. B}\ }\textbf {\bibinfo
  {volume} {106}},\ \bibinfo {pages} {045203} (\bibinfo {year}
  {2022})}\BibitemShut {NoStop}%
\bibitem [{\citenamefont {Seki}\ \emph {et~al.}(2010)\citenamefont {Seki},
  \citenamefont {Sugai}, \citenamefont {Hasegawa}, \citenamefont {Mitani},\
  and\ \citenamefont {Takanashi}}]{seki10}%
  \BibitemOpen
  \bibfield  {author} {\bibinfo {author} {\bibfnamefont {T.}~\bibnamefont
  {Seki}}, \bibinfo {author} {\bibfnamefont {I.}~\bibnamefont {Sugai}},
  \bibinfo {author} {\bibfnamefont {Y.}~\bibnamefont {Hasegawa}}, \bibinfo
  {author} {\bibfnamefont {S.}~\bibnamefont {Mitani}},\ and\ \bibinfo {author}
  {\bibfnamefont {K.}~\bibnamefont {Takanashi}},\ }\bibfield  {title} {\bibinfo
  {title} {{Spin Hall effect and Nernst effect in FePt/Au multi-terminal
  devices with different Au thicknesses}},\ }\href
  {https://doi.org/10.1016/j.ssc.2009.11.018} {\bibfield  {journal} {\bibinfo
  {journal} {Solid State Commun.}\ }\textbf {\bibinfo {volume} {150}},\
  \bibinfo {pages} {496} (\bibinfo {year} {2010})}\BibitemShut {NoStop}%
\bibitem [{\citenamefont {Meyer}\ \emph {et~al.}(2017)\citenamefont {Meyer},
  \citenamefont {Chen}, \citenamefont {Wimmer}, \citenamefont {Althammer},
  \citenamefont {Wimmer}, \citenamefont {Schlitz}, \citenamefont {Gepr\"{a}gs},
  \citenamefont {Huebl}, \citenamefont {K\"{o}dderitzsch}, \citenamefont
  {Ebert}, \citenamefont {Bauer}, \citenamefont {Gross},\ and\ \citenamefont
  {Goennenwein}}]{meyer17}%
  \BibitemOpen
  \bibfield  {author} {\bibinfo {author} {\bibfnamefont {S.}~\bibnamefont
  {Meyer}}, \bibinfo {author} {\bibfnamefont {Y.-T.}\ \bibnamefont {Chen}},
  \bibinfo {author} {\bibfnamefont {S.}~\bibnamefont {Wimmer}}, \bibinfo
  {author} {\bibfnamefont {M.}~\bibnamefont {Althammer}}, \bibinfo {author}
  {\bibfnamefont {T.}~\bibnamefont {Wimmer}}, \bibinfo {author} {\bibfnamefont
  {R.}~\bibnamefont {Schlitz}}, \bibinfo {author} {\bibfnamefont
  {S.}~\bibnamefont {Gepr\"{a}gs}}, \bibinfo {author} {\bibfnamefont
  {H.}~\bibnamefont {Huebl}}, \bibinfo {author} {\bibfnamefont
  {D.}~\bibnamefont {K\"{o}dderitzsch}}, \bibinfo {author} {\bibfnamefont
  {H.}~\bibnamefont {Ebert}}, \bibinfo {author} {\bibfnamefont {G.~E.~W.}\
  \bibnamefont {Bauer}}, \bibinfo {author} {\bibfnamefont {R.}~\bibnamefont
  {Gross}},\ and\ \bibinfo {author} {\bibfnamefont {S.~T.~B.}\ \bibnamefont
  {Goennenwein}},\ }\bibfield  {title} {\bibinfo {title} {{Observation of the
  spin Nernst effect}},\ }\href {https://doi.org/10.1038/nmat4964} {\bibfield
  {journal} {\bibinfo  {journal} {Nat. Mater.}\ }\textbf {\bibinfo {volume}
  {16}},\ \bibinfo {pages} {977} (\bibinfo {year} {2017})}\BibitemShut
  {NoStop}%
\bibitem [{\citenamefont {Sheng}\ \emph {et~al.}(2017)\citenamefont {Sheng},
  \citenamefont {Sakuraba}, \citenamefont {Lau}, \citenamefont {Takahashi},
  \citenamefont {Mitani},\ and\ \citenamefont {Hayashi}}]{sheng2017}%
  \BibitemOpen
  \bibfield  {author} {\bibinfo {author} {\bibfnamefont {P.}~\bibnamefont
  {Sheng}}, \bibinfo {author} {\bibfnamefont {Y.}~\bibnamefont {Sakuraba}},
  \bibinfo {author} {\bibfnamefont {Y.-C.}\ \bibnamefont {Lau}}, \bibinfo
  {author} {\bibfnamefont {S.}~\bibnamefont {Takahashi}}, \bibinfo {author}
  {\bibfnamefont {S.}~\bibnamefont {Mitani}},\ and\ \bibinfo {author}
  {\bibfnamefont {M.}~\bibnamefont {Hayashi}},\ }\bibfield  {title} {\bibinfo
  {title} {{The spin Nernst effect in tungsten}},\ }\href
  {https://doi.org/10.1126/sciadv.1701503} {\bibfield  {journal} {\bibinfo
  {journal} {Sci. Adv.}\ }\textbf {\bibinfo {volume} {3}},\ \bibinfo {pages}
  {e1701503} (\bibinfo {year} {2017})}\BibitemShut {NoStop}%
\bibitem [{\citenamefont {Kim}\ \emph {et~al.}(2017)\citenamefont {Kim},
  \citenamefont {Jeon}, \citenamefont {Choi}, \citenamefont {Lee},
  \citenamefont {Surabhi}, \citenamefont {Jeong}, \citenamefont {Lee},\ and\
  \citenamefont {Park}}]{kim2017}%
  \BibitemOpen
  \bibfield  {author} {\bibinfo {author} {\bibfnamefont {D.-J.}\ \bibnamefont
  {Kim}}, \bibinfo {author} {\bibfnamefont {C.-Y.}\ \bibnamefont {Jeon}},
  \bibinfo {author} {\bibfnamefont {J.-G.}\ \bibnamefont {Choi}}, \bibinfo
  {author} {\bibfnamefont {J.~W.}\ \bibnamefont {Lee}}, \bibinfo {author}
  {\bibfnamefont {S.}~\bibnamefont {Surabhi}}, \bibinfo {author} {\bibfnamefont
  {J.-R.}\ \bibnamefont {Jeong}}, \bibinfo {author} {\bibfnamefont {K.-J.}\
  \bibnamefont {Lee}},\ and\ \bibinfo {author} {\bibfnamefont {B.-G.}\
  \bibnamefont {Park}},\ }\bibfield  {title} {\bibinfo {title} {{Observation of
  transverse spin Nernst magnetoresistance induced by thermal spin current in
  ferromagnet/non-magnet bilayers}},\ }\href
  {https://doi.org/10.1038/s41467-017-01493-5} {\bibfield  {journal} {\bibinfo
  {journal} {Nat. Commun.}\ }\textbf {\bibinfo {volume} {8}},\ \bibinfo {pages}
  {1400} (\bibinfo {year} {2017})}\BibitemShut {NoStop}%
\bibitem [{\citenamefont {Ganichev}\ \emph {et~al.}(2012)\citenamefont
  {Ganichev}, \citenamefont {Trushin},\ and\ \citenamefont
  {Schliemann}}]{ganichev2012spin}%
  \BibitemOpen
  \bibfield  {author} {\bibinfo {author} {\bibfnamefont {S.~D.}\ \bibnamefont
  {Ganichev}}, \bibinfo {author} {\bibfnamefont {M.}~\bibnamefont {Trushin}},\
  and\ \bibinfo {author} {\bibfnamefont {J.}~\bibnamefont {Schliemann}},\
  }\href@noop {} {\bibinfo {title} {{Spin polarization by current in Handbook
  of Spin Transport and Magnetism, edited by E. Y, Tsymbal and I. Zutic}}}
  (\bibinfo {year} {Chapman \& Hall, Boca Raton, 2012})\BibitemShut {NoStop}%
\bibitem [{\citenamefont {Smirnov}\ and\ \citenamefont
  {Golub}(2017)}]{Hopping_spin}%
  \BibitemOpen
  \bibfield  {author} {\bibinfo {author} {\bibfnamefont {D.~S.}\ \bibnamefont
  {Smirnov}}\ and\ \bibinfo {author} {\bibfnamefont {L.~E.}\ \bibnamefont
  {Golub}},\ }\bibfield  {title} {\bibinfo {title} {{Electrical Spin
  Orientation, Spin-Galvanic, and Spin-Hall Effects in Disordered
  Two-Dimensional Systems}},\ }\href
  {https://doi.org/10.1103/PhysRevLett.118.116801} {\bibfield  {journal}
  {\bibinfo  {journal} {Phys. Rev. Lett.}\ }\textbf {\bibinfo {volume} {118}},\
  \bibinfo {pages} {116801} (\bibinfo {year} {2017})}\BibitemShut {NoStop}%
\bibitem [{\citenamefont {Glazov}(2018)}]{book_Glazov}%
  \BibitemOpen
  \bibfield  {author} {\bibinfo {author} {\bibfnamefont {M.~M.}\ \bibnamefont
  {Glazov}},\ }\href@noop {} {\emph {\bibinfo {title} {Electron and Nuclear
  Spin Dynamics in Semiconductor Nanostructures}}}\ (\bibinfo  {publisher}
  {Oxford University Press, Oxford},\ \bibinfo {year} {2018})\BibitemShut
  {NoStop}%
\bibitem [{\citenamefont {Perel'}\ \emph {et~al.}(2003)\citenamefont {Perel'},
  \citenamefont {Tarasenko}, \citenamefont {Yassievich}, \citenamefont
  {Ganichev}, \citenamefont {Bel'kov},\ and\ \citenamefont
  {Prettl}}]{perel:201304}%
  \BibitemOpen
  \bibfield  {author} {\bibinfo {author} {\bibfnamefont {V.~I.}\ \bibnamefont
  {Perel'}}, \bibinfo {author} {\bibfnamefont {S.~A.}\ \bibnamefont
  {Tarasenko}}, \bibinfo {author} {\bibfnamefont {I.~N.}\ \bibnamefont
  {Yassievich}}, \bibinfo {author} {\bibfnamefont {S.~D.}\ \bibnamefont
  {Ganichev}}, \bibinfo {author} {\bibfnamefont {V.~V.}\ \bibnamefont
  {Bel'kov}},\ and\ \bibinfo {author} {\bibfnamefont {W.}~\bibnamefont
  {Prettl}},\ }\bibfield  {title} {\bibinfo {title} {Spin-dependent tunneling
  through a symmetric semiconductor barrier},\ }\href
  {http://link.aps.org/abstract/PRB/v67/e201304} {\bibfield  {journal}
  {\bibinfo  {journal} {Phys. Rev. B}\ }\textbf {\bibinfo {volume} {67}},\
  \bibinfo {eid} {201304} (\bibinfo {year} {2003})}\BibitemShut {NoStop}%
\bibitem [{\citenamefont {Tarasenko}\ \emph {et~al.}(2004)\citenamefont
  {Tarasenko}, \citenamefont {Perel'},\ and\ \citenamefont
  {Yassievich}}]{tarasenko:056601}%
  \BibitemOpen
  \bibfield  {author} {\bibinfo {author} {\bibfnamefont {S.~A.}\ \bibnamefont
  {Tarasenko}}, \bibinfo {author} {\bibfnamefont {V.~I.}\ \bibnamefont
  {Perel'}},\ and\ \bibinfo {author} {\bibfnamefont {I.~N.}\ \bibnamefont
  {Yassievich}},\ }\bibfield  {title} {\bibinfo {title} {{In-Plane Electric
  Current Is Induced by Tunneling of Spin-Polarized Carriers}},\ }\href
  {http://link.aps.org/abstract/PRL/v93/e056601} {\bibfield  {journal}
  {\bibinfo  {journal} {Phys. Rev. Lett.}\ }\textbf {\bibinfo {volume} {93}},\
  \bibinfo {eid} {056601} (\bibinfo {year} {2004})}\BibitemShut {NoStop}%
\bibitem [{\citenamefont {Glazov}\ \emph {et~al.}(2005)\citenamefont {Glazov},
  \citenamefont {Alekseev}, \citenamefont {Odnoblyudov}, \citenamefont
  {Chistyakov}, \citenamefont {Tarasenko},\ and\ \citenamefont
  {Yassievich}}]{Tarasenko_Tunneling}%
  \BibitemOpen
  \bibfield  {author} {\bibinfo {author} {\bibfnamefont {M.~M.}\ \bibnamefont
  {Glazov}}, \bibinfo {author} {\bibfnamefont {P.~S.}\ \bibnamefont
  {Alekseev}}, \bibinfo {author} {\bibfnamefont {M.~A.}\ \bibnamefont
  {Odnoblyudov}}, \bibinfo {author} {\bibfnamefont {V.~M.}\ \bibnamefont
  {Chistyakov}}, \bibinfo {author} {\bibfnamefont {S.~A.}\ \bibnamefont
  {Tarasenko}},\ and\ \bibinfo {author} {\bibfnamefont {I.~N.}\ \bibnamefont
  {Yassievich}},\ }\bibfield  {title} {\bibinfo {title} {Spin-dependent
  resonant tunneling in symmetrical double-barrier structures},\ }\href
  {https://doi.org/10.1103/PhysRevB.71.155313} {\bibfield  {journal} {\bibinfo
  {journal} {Phys. Rev. B}\ }\textbf {\bibinfo {volume} {71}},\ \bibinfo
  {pages} {155313} (\bibinfo {year} {2005})}\BibitemShut {NoStop}%
\bibitem [{\citenamefont {Shumilin}\ \emph {et~al.}(2018)\citenamefont
  {Shumilin}, \citenamefont {Smirnov},\ and\ \citenamefont
  {Golub}}]{PhysRevB.98.155304}%
  \BibitemOpen
  \bibfield  {author} {\bibinfo {author} {\bibfnamefont {A.~V.}\ \bibnamefont
  {Shumilin}}, \bibinfo {author} {\bibfnamefont {D.~S.}\ \bibnamefont
  {Smirnov}},\ and\ \bibinfo {author} {\bibfnamefont {L.~E.}\ \bibnamefont
  {Golub}},\ }\bibfield  {title} {\bibinfo {title} {Spin-related phenomena in
  the two-dimensional hopping regime in magnetic field},\ }\href
  {https://doi.org/10.1103/PhysRevB.98.155304} {\bibfield  {journal} {\bibinfo
  {journal} {Phys. Rev. B}\ }\textbf {\bibinfo {volume} {98}},\ \bibinfo
  {pages} {155304} (\bibinfo {year} {2018})}\BibitemShut {NoStop}%
\bibitem [{\citenamefont {Mantsevich}\ and\ \citenamefont
  {Smirnov}(2022)}]{Mantsevich2022}%
  \BibitemOpen
  \bibfield  {author} {\bibinfo {author} {\bibfnamefont {V.~N.}\ \bibnamefont
  {Mantsevich}}\ and\ \bibinfo {author} {\bibfnamefont {D.~S.}\ \bibnamefont
  {Smirnov}},\ }\bibfield  {title} {\bibinfo {title} {Current-induced hole spin
  polarization in a quantum dot via a chiral quasi bound state},\ }\href
  {https://doi.org/10.1039/d1nh00685a} {\bibfield  {journal} {\bibinfo
  {journal} {Nanoscale Horiz.}\ }\textbf {\bibinfo {volume} {7}},\ \bibinfo
  {pages} {752} (\bibinfo {year} {2022})}\BibitemShut {NoStop}%
\bibitem [{\citenamefont {Mantsevich}\ and\ \citenamefont
  {Smirnov}(2023)}]{Mantsevich2023}%
  \BibitemOpen
  \bibfield  {author} {\bibinfo {author} {\bibfnamefont {V.~N.}\ \bibnamefont
  {Mantsevich}}\ and\ \bibinfo {author} {\bibfnamefont {D.~S.}\ \bibnamefont
  {Smirnov}},\ }\bibfield  {title} {\bibinfo {title} {Kondo enhancement of
  current-induced spin accumulation in a quantum dot},\ }\href
  {https://doi.org/10.1103/PhysRevB.108.035409} {\bibfield  {journal} {\bibinfo
   {journal} {Phys. Rev. B}\ }\textbf {\bibinfo {volume} {108}},\ \bibinfo
  {pages} {035409} (\bibinfo {year} {2023})}\BibitemShut {NoStop}%
\bibitem{exp1} \addDima{K. Kobayashi, H. Aikawa, A. Sano, S. Katsumoto, and Y. Iye, Fano resonance in a quantum wire with a side-coupled quantum dot, Phys. Rev. B \textbf{70}, 035319 (2004).}
\bibitem{exp2} \addDima{M. Sato, H. Aikawa, K. Kobayashi, S. Katsumoto, and Y. Iye, Observation of the Fano-Kondo Antiresonance in a Quantum Wire with a Side-Coupled Quantum Dot, Phys. Rev. Lett. \textbf{95}, 066801 (2005).}
\bibitem{exp3} \addDima{S. Katsumoto, M. Sato, H. Aikawa, and Y. Iye, Effect of localized spins in coherent transport through quantum dots, Physica E \textbf{34}, 36 (2006).}
\bibitem{review1} \addDima{S. Katsumoto, Coherence and spin effects in quantum dots, J. Phys.: Condens. Matter. \textbf{19}, 233201 (2007).}
\bibitem [{\citenamefont {Anderson}(1961)}]{PhysRev.124.41}%
  \BibitemOpen
  \bibfield  {author} {\bibinfo {author} {\bibfnamefont {P.~W.}\ \bibnamefont
  {Anderson}},\ }\bibfield  {title} {\bibinfo {title} {Localized magnetic
  states in metals},\ }\href {https://doi.org/10.1103/PhysRev.124.41}
  {\bibfield  {journal} {\bibinfo  {journal} {Phys. Rev.}\ }\textbf {\bibinfo
  {volume} {124}},\ \bibinfo {pages} {41} (\bibinfo {year} {1961})}\BibitemShut
  {NoStop}%
\bibitem{teor1} \addDima{P. A. Orellana, F. Dom\'{\i}nguez-Adame, I. G\'omez, and M. L. Ladr\'on de Guevara, Transport through a quantum wire with a side quantum-dot array, Phys. Rev. B \textbf{67}, 085321 (2003).}
\bibitem{teor2} \addDima{R. Franco, M. S. Figueira, and E. V. Anda, Fano resonance in electronic transport through a quantum wire with a side-coupled quantum dot: X-boson treatment, Phys. Rev. B \textbf{67}, 155301 (2003).}
\bibitem{teor3} \addDima{A. C. Seridonio, M. Yoshida, and L. N. Oliveira, Universal zero-bias conductance through a quantum wire side-coupled to a quantum dot, Phys. Rev. B \textbf{80}, 235318 (2009).}
\bibitem [{\citenamefont {Rammer}\ and\ \citenamefont
  {Smith}(1986)}]{RevModPhys.58.323}%
  \BibitemOpen
  \bibfield  {author} {\bibinfo {author} {\bibfnamefont {J.}~\bibnamefont
  {Rammer}}\ and\ \bibinfo {author} {\bibfnamefont {H.}~\bibnamefont {Smith}},\
  }\bibfield  {title} {\bibinfo {title} {Quantum field-theoretical methods in
  transport theory of metals},\ }\href
  {https://doi.org/10.1103/RevModPhys.58.323} {\bibfield  {journal} {\bibinfo
  {journal} {Rev. Mod. Phys.}\ }\textbf {\bibinfo {volume} {58}},\ \bibinfo
  {pages} {323} (\bibinfo {year} {1986})}\BibitemShut {NoStop}%
\bibitem [{\citenamefont {Stefanucci}\ and\ \citenamefont {van
  Leeuwen}(2013)}]{stefanucci_vanleeuwen_2013}%
  \BibitemOpen
  \bibfield  {author} {\bibinfo {author} {\bibfnamefont {G.}~\bibnamefont
  {Stefanucci}}\ and\ \bibinfo {author} {\bibfnamefont {R.}~\bibnamefont {van
  Leeuwen}},\ }\href {https://doi.org/10.1017/CBO9781139023979} {\emph
  {\bibinfo {title} {{Nonequilibrium Many-Body Theory of Quantum Systems: A
  Modern Introduction}}}}\ (\bibinfo  {publisher} {Cambridge University
  Press},\ \bibinfo {year} {2013})\BibitemShut {NoStop}%
\bibitem [{\citenamefont {Arseev}(2015)}]{Arseev_2015}%
  \BibitemOpen
  \bibfield  {author} {\bibinfo {author} {\bibfnamefont {P.~I.}\ \bibnamefont
  {Arseev}},\ }\bibfield  {title} {\bibinfo {title} {On the nonequilibrium
  diagram technique: derivation, some features, and applications},\ }\href
  {https://doi.org/10.3367/ufne.0185.201512b.1271} {\bibfield  {journal}
  {\bibinfo  {journal} {Phys. Usp.}\ }\textbf {\bibinfo {volume} {58}},\
  \bibinfo {pages} {1159} (\bibinfo {year} {2015})}\BibitemShut {NoStop}%
\bibitem [{\citenamefont {Hubbard}\ and\ \citenamefont
  {Flowers}(1963)}]{doi:10.1098/rspa.1963.0204}%
  \BibitemOpen
  \bibfield  {author} {\bibinfo {author} {\bibfnamefont {J.}~\bibnamefont
  {Hubbard}}\ and\ \bibinfo {author} {\bibfnamefont {B.~H.}\ \bibnamefont
  {Flowers}},\ }\bibfield  {title} {\bibinfo {title} {Electron correlations in
  narrow energy bands},\ }\href {https://doi.org/10.1098/rspa.1963.0204}
  {\bibfield  {journal} {\bibinfo  {journal} {J. Proc. Roy. Soc. A}\ }\textbf
  {\bibinfo {volume} {276}},\ \bibinfo {pages} {238} (\bibinfo {year}
  {1963})}\BibitemShut {NoStop}%
\bibitem [{\citenamefont {Haug}\ and\ \citenamefont
  {Jauho}(2008)}]{haug2008quantum}%
  \BibitemOpen
  \bibfield  {author} {\bibinfo {author} {\bibfnamefont {H.}~\bibnamefont
  {Haug}}\ and\ \bibinfo {author} {\bibfnamefont {A.-P.}\ \bibnamefont
  {Jauho}},\ }\href@noop {} {\emph {\bibinfo {title} {Quantum kinetics in
  transport and optics of semiconductors}}},\ Vol.~\bibinfo {volume} {2}\
  (\bibinfo  {publisher} {Springer},\ \bibinfo {year} {2008})\BibitemShut
  {NoStop}%
\bibitem [{\citenamefont {Meir}\ \emph {et~al.}(1993)\citenamefont {Meir},
  \citenamefont {Wingreen},\ and\ \citenamefont {Lee}}]{PhysRevLett.70.2601}%
  \BibitemOpen
  \bibfield  {author} {\bibinfo {author} {\bibfnamefont {Y.}~\bibnamefont
  {Meir}}, \bibinfo {author} {\bibfnamefont {N.~S.}\ \bibnamefont {Wingreen}},\
  and\ \bibinfo {author} {\bibfnamefont {P.~A.}\ \bibnamefont {Lee}},\
  }\bibfield  {title} {\bibinfo {title} {Low-temperature transport through a
  quantum dot: The anderson model out of equilibrium},\ }\href
  {https://doi.org/10.1103/PhysRevLett.70.2601} {\bibfield  {journal} {\bibinfo
   {journal} {Phys. Rev. Lett.}\ }\textbf {\bibinfo {volume} {70}},\ \bibinfo
 {pages} {2601} (\bibinfo {year} {1993})}\BibitemShut {NoStop}%
\bibitem{Pols} R. {\'S}wirkowicz, M. Wilczy{\'n}ski, J. Barna{\'s}, Spin-polarized transport through a single-level quantum dot in the Kondo regime, J. Phys.: Condens. Matter \textbf{18}, 2291 (2006).
\bibitem [{\citenamefont {Mahan}(2013)}]{mahan2013many}%
  \BibitemOpen
  \bibfield  {author} {\bibinfo {author} {\bibfnamefont {G.~D.}\ \bibnamefont
  {Mahan}},\ }\href@noop {} {\emph {\bibinfo {title} {Many-particle physics}}}\
  (\bibinfo  {publisher} {Springer Science \& Business Media},\ \bibinfo {year}
  {2013})\BibitemShut {NoStop}%
\bibitem [{\citenamefont {Lacroix}(1981)}]{Lacroix_1981}%
  \BibitemOpen
  \bibfield  {author} {\bibinfo {author} {\bibfnamefont {C.}~\bibnamefont
  {Lacroix}},\ }\bibfield  {title} {\bibinfo {title} {Density of states for the
  anderson model},\ }\href {https://doi.org/10.1088/0305-4608/11/11/020}
  {\bibfield  {journal} {\bibinfo  {journal} {J. Phys. F}\ }\textbf {\bibinfo
  {volume} {11}},\ \bibinfo {pages} {2389} (\bibinfo {year}
  {1981})}\BibitemShut {NoStop}%
\bibitem [{\citenamefont {Arseyev}\ \emph {et~al.}(2012)\citenamefont
  {Arseyev}, \citenamefont {Maslova},\ and\ \citenamefont
  {Mantsevich}}]{Arseyev2012}%
  \BibitemOpen
  \bibfield  {author} {\bibinfo {author} {\bibfnamefont {P.~I.}\ \bibnamefont
  {Arseyev}}, \bibinfo {author} {\bibfnamefont {N.~S.}\ \bibnamefont
  {Maslova}},\ and\ \bibinfo {author} {\bibfnamefont {V.~N.}\ \bibnamefont
  {Mantsevich}},\ }\bibfield  {title} {\bibinfo {title} {Coulomb correlations
  effects on localized charge relaxation in the coupled quantum dots},\
  }\bibfield  {journal} {\bibinfo  {journal} {The European Physical Journal B}\
  }\textbf {\bibinfo {volume} {85}},\ \href
  {https://doi.org/10.1140/epjb/e2012-20948-x} {10.1140/epjb/e2012-20948-x}
  (\bibinfo {year} {2012})\BibitemShut {NoStop}%
\bibitem [{\citenamefont {Mantsevich}\ \emph {et~al.}(2013)\citenamefont
  {Mantsevich}, \citenamefont {Maslova},\ and\ \citenamefont
  {Arseyev}}]{Mantsevich2013}%
  \BibitemOpen
  \bibfield  {author} {\bibinfo {author} {\bibfnamefont {V.}~\bibnamefont
  {Mantsevich}}, \bibinfo {author} {\bibfnamefont {N.}~\bibnamefont
  {Maslova}},\ and\ \bibinfo {author} {\bibfnamefont {P.}~\bibnamefont
  {Arseyev}},\ }\bibfield  {title} {\bibinfo {title} {Charge trapping in the
  system of interacting quantum dots},\ }\href
  {https://doi.org/10.1016/j.ssc.2013.06.017} {\bibfield  {journal} {\bibinfo
  {journal} {Solid State Communications}\ }\textbf {\bibinfo {volume} {168}},\
  \bibinfo {pages} {36–41} (\bibinfo {year} {2013})}\BibitemShut {NoStop}%
\bibitem [{\citenamefont {Mantsevich}\ and\ \citenamefont
  {Maslova}(2010)}]{Mantsevich2010}%
  \BibitemOpen
  \bibfield  {author} {\bibinfo {author} {\bibfnamefont {V.}~\bibnamefont
  {Mantsevich}}\ and\ \bibinfo {author} {\bibfnamefont {N.}~\bibnamefont
  {Maslova}},\ }\bibfield  {title} {\bibinfo {title} {Different behaviour of
  local tunneling conductivity for deep and shallow impurities due to coulomb
  interaction},\ }\href {https://doi.org/10.1016/j.ssc.2010.07.051} {\bibfield
  {journal} {\bibinfo  {journal} {Solid State Communications}\ }\textbf
  {\bibinfo {volume} {150}},\ \bibinfo {pages} {2072–2075} (\bibinfo {year}
  {2010})}\BibitemShut {NoStop}%
\bibitem [{\citenamefont {Arseev}\ \emph {et~al.}(2012)\citenamefont {Arseev},
  \citenamefont {Maslova},\ and\ \citenamefont {Mantsevich}}]{Arseev2012}%
  \BibitemOpen
  \bibfield  {author} {\bibinfo {author} {\bibfnamefont {P.~I.}\ \bibnamefont
  {Arseev}}, \bibinfo {author} {\bibfnamefont {N.~S.}\ \bibnamefont
  {Maslova}},\ and\ \bibinfo {author} {\bibfnamefont {V.~N.}\ \bibnamefont
  {Mantsevich}},\ }\bibfield  {title} {\bibinfo {title} {The effect of coulomb
  correlations on the nonequilibrium charge redistribution tuned by the
  tunneling current},\ }\href {https://doi.org/10.1134/s1063776112060027}
  {\bibfield  {journal} {\bibinfo  {journal} {Journal of Experimental and
  Theoretical Physics}\ }\textbf {\bibinfo {volume} {115}},\ \bibinfo {pages}
  {141–153} (\bibinfo {year} {2012})}\BibitemShut {NoStop}%
\bibitem [{\citenamefont {Arseyev}\ \emph {et~al.}(2011)\citenamefont
  {Arseyev}, \citenamefont {Maslova},\ and\ \citenamefont
  {Mantsevich}}]{Arseyev2011}%
  \BibitemOpen
  \bibfield  {author} {\bibinfo {author} {\bibfnamefont {P.~I.}\ \bibnamefont
  {Arseyev}}, \bibinfo {author} {\bibfnamefont {N.~S.}\ \bibnamefont
  {Maslova}},\ and\ \bibinfo {author} {\bibfnamefont {V.~N.}\ \bibnamefont
  {Mantsevich}},\ }\bibfield  {title} {\bibinfo {title} {Correlation induced
  switching of the local spatial charge distribution in a two-level system},\
  }\href {https://doi.org/10.1134/s0021364011170048} {\bibfield  {journal}
  {\bibinfo  {journal} {JETP Letters}\ }\textbf {\bibinfo {volume} {94}},\
  \bibinfo {pages} {390–396} (\bibinfo {year} {2011})}\BibitemShut {NoStop}%
\bibitem [{\citenamefont {Niu}\ \emph {et~al.}(1999)\citenamefont {Niu},
  \citenamefont {Lin},\ and\ \citenamefont {Lin}}]{Niu_1999}%
  \BibitemOpen
  \bibfield  {author} {\bibinfo {author} {\bibfnamefont {C.}~\bibnamefont
  {Niu}}, \bibinfo {author} {\bibfnamefont {D.~L.}\ \bibnamefont {Lin}},\ and\
  \bibinfo {author} {\bibfnamefont {T.-H.}\ \bibnamefont {Lin}},\ }\bibfield
  {title} {\bibinfo {title} {{Equation of motion for nonequilibrium Green
  functions}},\ }\href {https://doi.org/10.1088/0953-8984/11/6/015} {\bibfield
  {journal} {\bibinfo  {journal} {J. Phys.: Cond. Mat.}\ }\textbf {\bibinfo
  {volume} {11}},\ \bibinfo {pages} {1511} (\bibinfo {year}
  {1999})}\BibitemShut {NoStop}%
\end{thebibliography}
\end{document}